# New energy conversion system based on charge-exchange and inner-shell electron transitions

*Tianrui Li[a], Yi Jiang[a], Chen Zhao[a], Bingsheng Tu[b], Peining Chen[a], Jiajun Qin[a], Huisheng Peng[a]\**

[a] State Key Laboratory of Molecular Engineering of Polymers, Department of Macromolecular Science, Laboratory of Advanced Materials, Institute of Fiber Materials and Devices, Fudan University, Shanghai 200433, China.

E-mail: penghs@fudan.edu.cn.

[b] Key Laboratory of Nuclear Physics and Ion-Beam Application, Institute of Modern Physics, Fudan University, Shanghai 200433, China.

ABSTRACT: The rapidly growing demand for compact, high-energy power sources has outpaced the capabilities of conventional electrochemical systems that rely on outer-shell redox reactions. In this work, we present a new energy platform that utilizes inner-shell electron transitions that are previously inaccessible due to their high energy thresholds. By leveraging charge exchange processes between bare argon ions ($Ar^{18+}$) and neutral helium atoms, we provide clear evidence for the emission of soft X-ray and extreme-ultraviolet photons across a broad spectra range, resulting from inner-shell electron capture and cascade de-excitation. This strategy overcomes the limitations of radiative recombination by enhancing photon energy utilization through broader emission profiles more compatible with practical energy converters. Our design of a helium-filled chamber design enables precise control of output via pressure tuning, achieving a remarkable radiation power density of $6.29 \times 10^8$ W L$^{-1}$ and an unprecedented energy density of $2.64 \times 10^6$ Wh kg$^{-1}$. These results may provide a new and effective paradigm for energy conversion systems with ultra-high power and energy densities based on inner-shell electrons.

## 1. Introduction

The escalating power and energy demand of modern technologies, ranging from electric vehicles to aerospace systems, are rapidly exceeding the capabilities of conventional electrochemical storage systems, such as lithium-ion batteries, which rely on outer-shell redox reactions. In contrast, inner-shell electron transitions release orders of magnitude more energy. A single K-shell binding energy in Ar is around 3.2 keV[1], nearly a thousand times the energy per electron transferred in LiCoO₂ redox processes in batteries[2]. Therefore, harnessing these inner-shell processes could present the next-generation energy solution for extremely high power and energy densities.

Despite of the mentioned potential above, the direct utilization of inner-shell electrons through conventional electrochemical or photoexcitation techniques remains infeasible. The strong nuclear Coulomb attraction result in extremely high energy barriers, thus preventing the direct utilization of inner-shell electrons[3-5]. A promising alternative exploits the inherently high ionization potential of highly charged ions. Radiative recombination tends to produce hard X-rays via direct relaxation to the ground state, limiting photon utility[6]. In contrast, charge exchange favors population of high-n states followed by cascade decay, yielding more accessible and usable photons. These high-energy photons may be hard to couple with practical photoelectric converters. Charge exchange, in which electrons into high-n levels, leading to cascade decay and the emission of photons across a broader spectrum. This includes soft X-rays and ultraviolet light, which are more efficiently convertible for practical energy applications [7-10].

Here we report a new energy conversion system with both high power and energy densities based on charge exchange between bare Argon (Ar) ions and helium gas. In these processes, each Ar18+ captures an electron into a high-n orbital, followed by a de-excitation cascade that predominantly emits soft X-ray and extreme-ultraviolet (EUV) photons (Figure 1a). A helium-filled collision chamber and a photoelectric converter are coupled as the energy collection module (Figure 1b). In this design, helium acts both as an electron donor, eliminating a separate electron gun, and as a protective medium that prevents damage to the chamber and converter[11-13]. The output power can be controlled by tuning the helium pressure. Under a He pressure of 20 atm, the energy conversion system produces a radiation power density of $6.29 \times 10^8$ W L$^{-1}$ and an energy density of $2.64 \times 10^6$ Wh kg$^{-1}$.

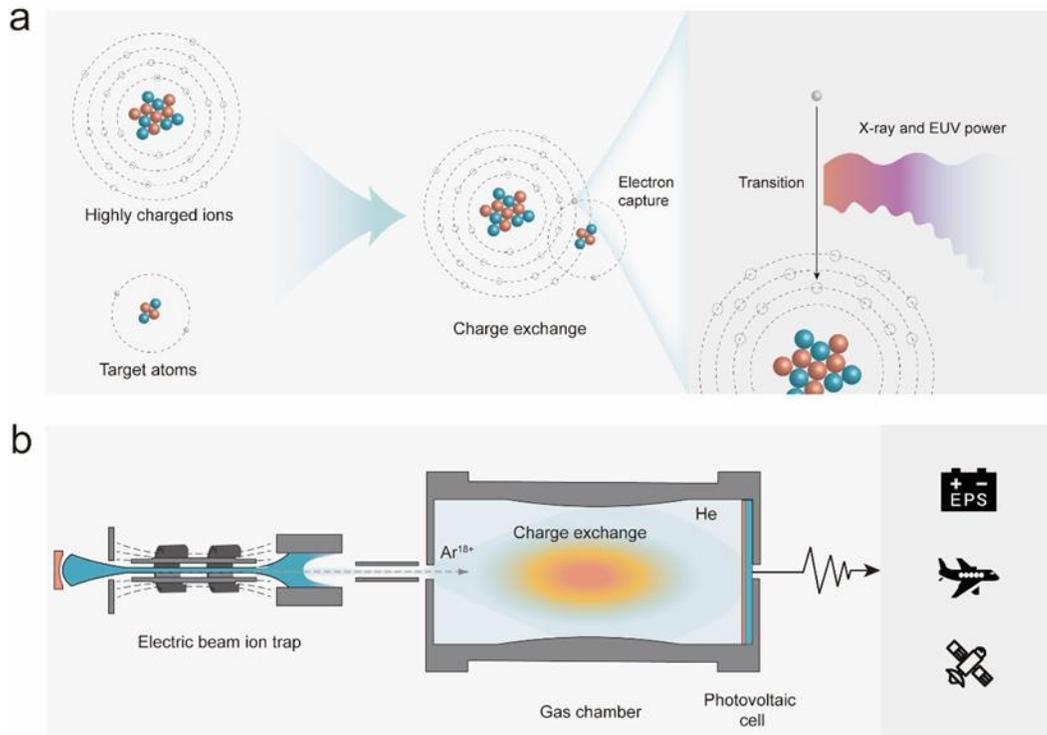

**Figure 1.** a) Schematic diagram of charge exchange and corresponding photon emission.

b) The overall design for the energy conversion system.

## 2. Methods

*Atomic structure and charge exchange calculations*: All the caculations were carried out with the Flexible Atomic Code.

*Data collection and collation*: All the caculated data was collected and collated by designed Python codes based on scientific formulas (Supporting Information).

## 3. Results and discussion

All atomic structure and charge exchange calculations are carried out with the Flexible Atomic Code (FAC). FAC is an integrated toolkit for computing a large number of atomic processes, including photoionization, autoionization, radiative and dielectronic recombination, collisional excitation, and charge exchange[14]. FAC evaluates charge-exchange (CX) cross sections for bare and highly charged ions via Landau-Zener model. For collision energies up to 200 eV u⁻¹, FAC solves the full quantum-mechanical coupled-channels equations, while for the above threshold it employs hydrogenic approximations for bound-state wavefunctions.

In the multi-channel Landau-Zener collision model implemented by FAC[15], the energy release of charge exchange includes two steps: non-radiative electron capture and radiative deexcitation[16]. After the collision occurs, the highly charged Ar ion captures an electron from a He atom into a high-lying orbital via Coulomb interaction. Subsequently, the captured electron rapidly transits to a low energy level, emitting a certain energy photon:

$$Ar^{18+} + He \rightarrow Ar^{17+}_{nl} + He^+ \quad (1)$$

$$Ar^{17+}_{nl} \rightarrow Ar^{17+}_{1s} + h\nu \quad (2)$$

Here, $Ar^{17+}_{nl}$ represents the excited $Ar^{17+}$ ion immediately following electron capture. $Ar^{17+}_{1s}$ represents the ground state. The orbital quantum number of ground state $Ar^{17+}$ is definite, whereas the orbitals of excited $Ar^{17+}$ are uncertain at multiple energy levels, with the capture probability into each level set by the corresponding charge exchange cross sections. De-excitation to the ground configuration then proceeds via a cascade of radiative transitions through multiple intermediate levels[17], and each step emits a photon with energy equals to the corresponding level spacing.

In the intermediate to high collision regime (⩾200 eV), the Landau-Zener capture model predicts the principal quantum number (n) of the captured orbital using the $q^{0.75}$ scaling rule [18].

For $Ar^{18+}$, n = 8.739, which sets the predicted principal quantum number of 8 or 9.

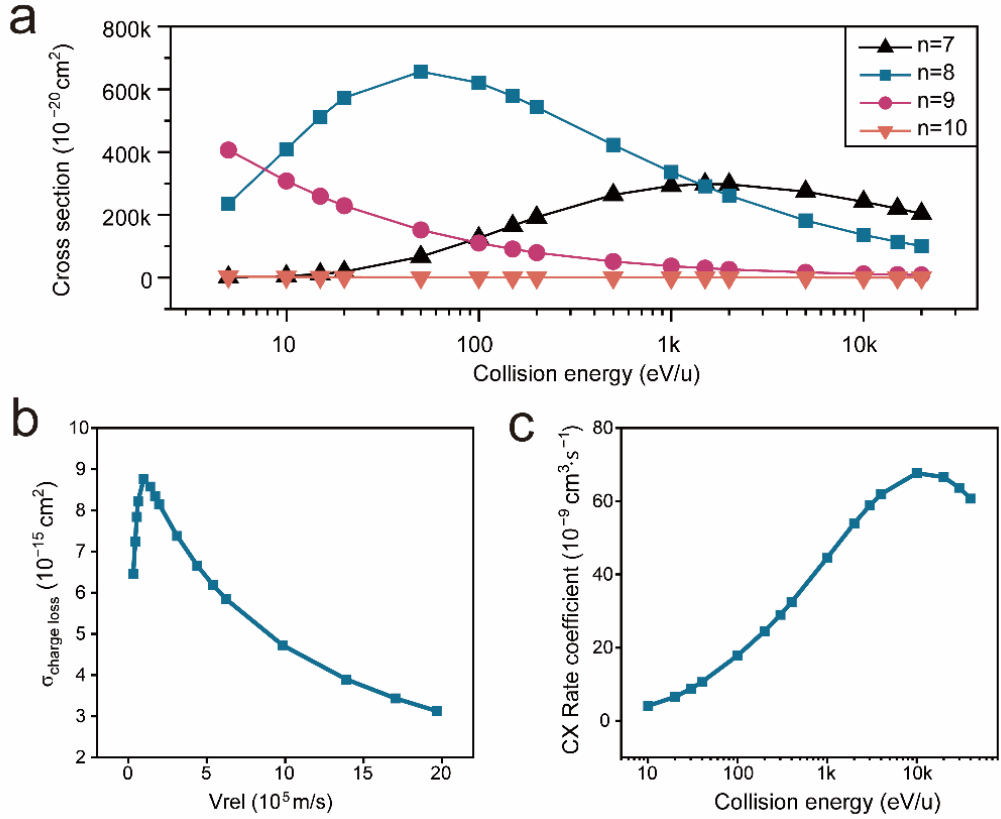

**Figure 2.** Collision energy dependence of the charge exchange process. a) Charge exchange cross sections for captures into n=7,8,9,10 as a function of collision energy (5–20 keV u$^{-1}$). b) Total charge loss cross section distribution under the particle velocity grid (0–2×10$^6$ m s$^{-1}$). c) Charge exchange rate coefficient against collision energy (5–20 keV u$^{-1}$).

Guided by the $q^{0.75}$ prediction, we use the FAC to calculate charge exchange cross sections for Ar$^{18+}$ and He across the electron capture to $n$=5–12 and collision energies of 5–20 keV u$^{-1}$. The charge exchange cross section quantifies the effective area associated with a single charge transfer event, representing the likelihood of charge exchange during particle collisions[19], and constitutes a key parameter in this analysis. A comparison of cross sections for different principal quantum numbers shows that, at intermediate to high collision energies, the capture into $n$=7–10 dominates. For $n$=7,8, the cross sections exhibit distinct peaks at specific collision energies; whereas for $n$=9,10, the cross sections decrease with increasing collision energy(Figure 2a). When the principal quantum number of $n$ is equal to or higher than 11, the charge exchange cross section drops by orders of magnitude ($10^{-22}$ cm$^2$ at $n$=11 and ~$10^{-35}$ cm$^2$ at $n$=12) (Figure S1), owing to the mismatch between the binding energies of these high states and the He ionization potential. A similar suppression is also observed for n≤6. Therefore, the subsequent calculations will primarily focus on the range of $n$=7–10.

To investigate the contribution of each orbital, we have calculated the collision energy-dependent cross sections for various orbitals at $n$=7,8,9,10 using FAC (Figure S2). Each orbital is described by its angular and spin quantum number. At $n$=7,8, the cross sections for orbitals beyond the f-orbital decrease significantly, accompanied by a shift in the peak positions. In contrast, for $n$=9,10, all orbitals decay smoothly and monotonically with collision energy, indicating that the capture into these higher shells is governed primarily by the increasing relative velocity. A portion of the captured electrons populate high-n orbitals beyond the d shell, which favor cascade decay pathways and lead to the emission of photons with greater practical utility.

To further reveal the charge exchange process, we have evaluated the total charge loss cross section of He atoms acting as target species colliding with Ar$^{18+}$. The total charge loss cross section is defined as the sum of the cross sections of charge exchange process that result in a net loss of charge from the incident ion[20]. In the single-electron capture process between an Ar$^{18+}$ ion and a He atom, the total charge loss section of the He atom reflects the electron transfer and associated energy exchange during the collision. Specifically, when an Ar$^{18+}$ ion collides with a He atom, it captures one electron from the He atom, resulting in the formation of He$^+$. This charge loss of the He atom indicates the occurrence of electron transfer in the interaction. As shown in Figure 2b, The single-electron capture cross section typically exhibits a peak as a function of collision energy due to a balance between interaction time and electronic coupling strength. At low energies, the prolonged interaction time is insufficient for efficient electron transfer due to weak nonadiabatic coupling. As the energy increases, described by the Landau-Zener model, the coupling between molecular states becomes more favorable, enabling more probable electron capture. This results in a rise in the cross section up to an optimal energy range. At higher energies, however, the relative velocity becomes too large, reducing the effective interaction time and suppressing the transition probability, thus causing the cross section to decrease. This peak behavior is a general feature in intermediate-energy ion-atom collisions and reflects the interplay between quantum dynamics and classical interaction time scales[21].

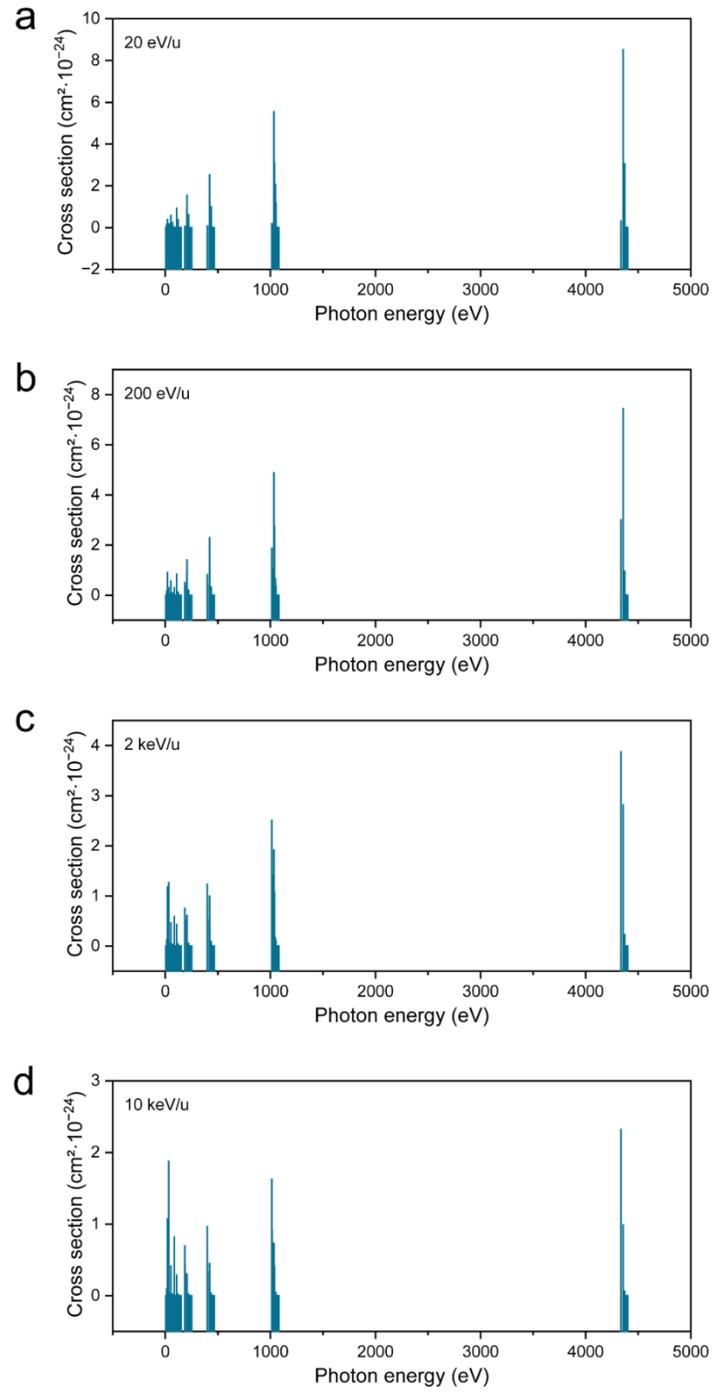

**Figure 3.** Photon energy distributions at collision energies of a) 20 eV u$^{-1}$, b) 200 eV u$^{-1}$, c) 2 keV u$^{-1}$, d) 10 keV u$^{-1}$.

By multiplying each charge-exchange cross section by the ratio for every possible radiative decay (Supporting Information), we may obtain the photon energy distributions at various collision energies (Figure 3). Four collision energies are set, including 20 eV u$^{-1}$, 200 eV·u$^{-1}$, 2 keV u$^{-1}$, and 10 keV u$^{-1}$. In each case, the emission can be divided into three main spectral regions: a dominant soft X-ray region (0.1–10 keV), a weaker dominant EUV band (10–124 eV), and a visible band (1.6–3 eV). The spectral profile reflects the electronic decay processes following the electron capture. Direct transition into deep inner shells (L shell) produces soft X-ray photons, while other electrons cascade through two or more intermediate states, producing a chain of lower-energy photons in the EUV and even visible bands. With the direct conversion of soft X-rays into voltage, or first with the conversion of X-rays into visible light and then the absorption, these forms of radiation energy can be effectively converted into usable electrical power by photovoltaic cells or hybrid devices based on SiC[22] or GaAs[23] semiconductors.

In a He-rich environment, we simulate the time-dependent evolution of the portion of Ar$^{18+}$ over a range of collision energies (Figure 4a). The density of He atoms is set to 2.69×10$^{18}$ cm$^{-3}$, corresponding to a He gas pressure of 0.1 atm in 1 cm³ volume. Within the single-electron capture Landau-Zener model, the Ar$^{18+}$ ion captures one electron from He atom to become Ar$^{17+}$, leading to a gradual decrease of average charge state. The rate of this change can be determined by the charge exchange cross section, He pressure and relative velocity. Consequently, the mean Ar charge state can be obtained. At a collision energy of 10 keV u$^{-1}$ and a He pressure at 0.1 atm, the charge exchange rate coefficient reaches its maximum (Figure 2c). Within 2×10$^{-8}$ s, Ar$^{18+}$ ions undergo complete single-electron charge exchange. Such a short reaction time indicates the significant potential for high-power output. With the continuous injection of highly charged ions into the system, sustained energy release can be achieved. Furthermore, the distance $L_c$ required for an ion to complete single-electron charge exchange process can be estimated using the charge exchange completion time $t_c$ and the relative velocity $v_{rel}$:

$$L_c = t_c \cdot v_{rel} \tag{3}$$

It is found that even at the lowest $v_{rel}$ of 20 eV u$^{-1}$, the estimated charge exchange path length $L_c$ is as short as 0.173 cm, which significantly favors the compact design and miniaturization of the proposed energy conversion system.

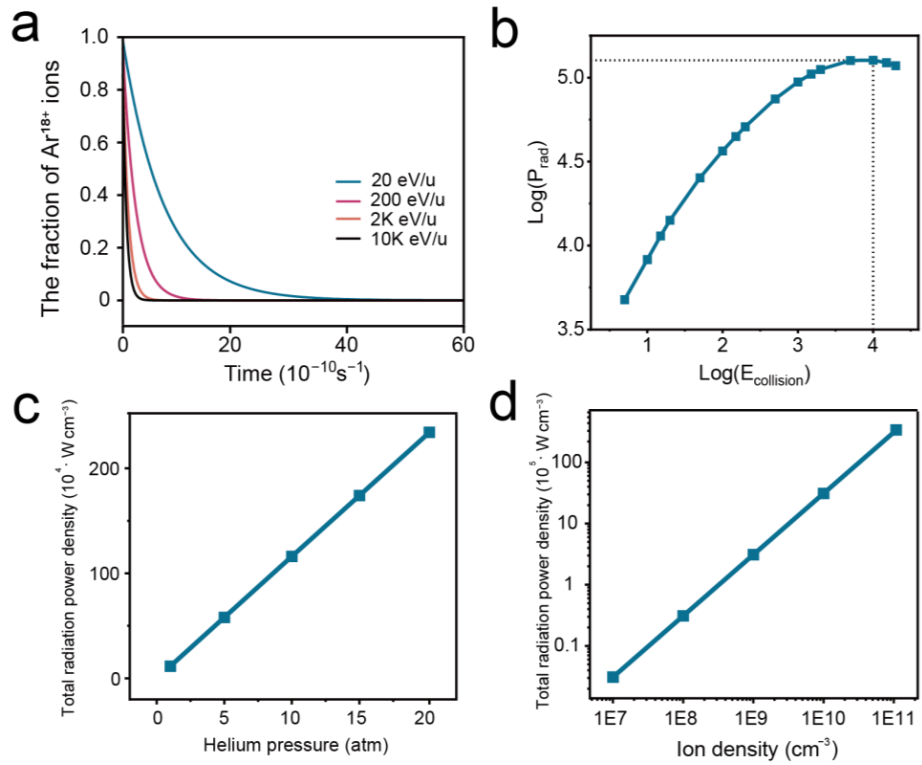

**Figure 4**. a) The time evolution of the Ar$^{18+}$ fraction under different collision energies (He density = 0.1 atm cm$^{-3}$). b) Double logarithmic plot illustrating the dependence of radiated power $P_{rad}$ on collision energy $E_{collision}$. c) Dependence of total radiation power on helium gas pressure. d) Dependence of total radiation power on Ar$^{18+}$ ion density.

Building upon the preceding data analysis, we further evaluated the radiation power and energy density of the energy conversion system with highly charged Ar ions. The radiation power density at a given collision energy is determined by the transition cross section, ion and target atom densities, and the interaction volume. By integrating the all photon energies released in charge exchange induced transitions, we may calculate the total radiation power under various collision energies as Figure 4b. The He pressure is set to 0.1 atm, and the ion beam density is set to $7.5\times10^{10}$ cm$^{-3}$, consistent with practical modern scientific instrument conditions[24]. The results reveal a peak in total radiation power at a collision energy of 10 keV u$^{-1}$. At higher energies, the decreasing charge exchange cross section leads to a decline in radiation power. The total radiation power exhibits a linear dependence on both He atom and ion density (Figure 4c, 4d). Under 10 keV u$^{-1}$ collision energy, increasing the He atom density and the ion density from $10^7$ to $10^{11}$ cm$^{-3}$ results in a maximum power density of $6.29\times10^8$ W L$^{-1}$. Based on the calculated characteristic time of the full charge exchange process, the energy density is found to be $2.64\times106$ Wh kg$^{-1}$.

## 4. Conclusion

In summary, we have proposed and theoretically validated a novel energy conversion system that exploits electron charge exchange and subsequent radiative cascades in bare Ar$^{18+}$ ions colliding with He. He serves both as the electron donor, eliminating the need for a separate electron gun, and a buffer gas that protects the chamber walls and photoelectric converter parts from ion damage. Through charge exchange process between highly charged Ar$^{18+}$ ions and helium atoms, the system produces a broad spectrum of soft X-ray and EUV photons, which can be easier harvested by existing photoconversion technologies. Under optimized conditions, the system achieves a peak radiation power density of $6.29\times10^8$ W L$^{-1}$ and energy densities of $2.64\times10^6$ Wh kg$^{-1}$, far surpassing the conventional electrochemical storage. This work thus offers a promising pathway toward extremely high energy density power sources. In future studies, selecting alternative projectile ions and target gases to shift emission into the visible band could further enhance photoelectric conversion efficiency and broaden practical applications.

## ASSOCIATED CONTENT

The Supporting Information includes detailed calculation methods, physical theories and supporting figures.

The following file is available free of charge.

supporting information (DOCX)

## AUTHOR INFORMATION

**Author Contributions**

Huisheng Peng directed all aspects of the project. Tianrui Li conceived and carried out all the calculations and wrote the paper. Yi Jiang and Chen Zhao assisted with the schematic diagram. Chen Zhao and Bingsheng Tu provided assistance with manuscript preparation.

**Funding Sources**

This article is funded by MOST (2022YFA1203001, 2022YFA1203002), NSFC (T2321003, 22335003, 52122310, 22075050), and STCSM (21511104900).

## ACKNOWLEDGMENT

We acknowledge Prof. Ke Yao for their assistance with the FAC and discussion.

ABBREVIATIONS

EUV, extreme-ultraviolet; FAC, Flexible Atomic Code.